# A Process to Facilitate Automated Automotive Cybersecurity Testing

Stefan Marksteiner
AVL List
Email: stefan.marksteiner@avl.com

Nadja Marko
Virtual Vehicle Resarch
Email: nadja.marko@v2c2.at

Andre Smulders
TNO
Email: andre.smulders@tno.nl

Stelios Karagiannis
Beyond Vision
Email: stelios.karagiannis@beyond-vision.pt

Florian Stahl
AVL Software & Functions
Email: florian.stahl@avl.com

Hayk Hamazaryan
ZF Friedrichshafen
Email: hayk.hamazaryan@zf.com

Rupert Schlick
Austrian Institute of Technology
Email: rupert.schlick@ait.ac.at

Stefan Kraxberger
SecInto
Email: stefan.kraxberger@secinto.com

Alexandr Vasenev
Joint Innovation Centre ESI (TNO)
Email: alexandr.vasenev@tno.nl

*Abstract*—Modern vehicles become increasingly digitalized with advanced information technology-based solutions like advanced driving assistance systems and vehicle-to-x communications. These systems are complex and interconnected. Rising complexity and increasing outside exposure has created a steadily rising demand for more cyber-secure systems. Thus, also standardization bodies and regulators issued standards and regulations to prescribe more secure development processes. This security, however, also has to be validated and verified. In order to keep pace with the need for more thorough, quicker and comparable testing, today's generally manual testing processes have to be structured and optimized. Based on existing and emerging standards for cybersecurity engineering, this paper therefore outlines a structured testing process for verifying and validating automotive cybersecurity, for which there is no standardized method so far. Despite presenting a commonly structured framework, the process is flexible in order to allow implementers to utilize their own, accustomed toolsets.

## I. INTRODUCTION

The rising complexity of modern automotive systems make it increasingly difficult to assure their cybersecurity. This is especially true due to the utilization of advanced driving assistance systems (ADAS) and autonomous driving (AD) and the exposure to the outside by to vehicle-to-x (V2X) functions. This usage of new technology is likely to accelerate even more; also market-leading manufacturers are beginning to equip their most-selling model with V2X off-the-shelf [1]. These developments facilitate cybersecurity incidents (e.g. [2], [3]). Furthermore, an exponential rise of events and an accel-eratingly adverse ratio between criminal activities versus benevolent security research results can be observed [4]. This has also been recognized by standardization bodies – currently, the most important standardization effort is ISO/SAE DIS 21434 [5]. Also, regulators begin to take cybersecurity considerations into account; the most prominent example being a recent regulation by the United Nations [6]. The rising incidents and the regulators' requirements demand a substantially higher amount of cybersecurity engineering and testing of automotive systems. This also creates the need for higher efficiency which makes a standardized method of automotive cybersecurity testing necessary. Currently, automotive cybersecurity testing is mostly not holistic, unstructured, non-reproducible and more art than crafts. An approach to giving a standardized and industrial-grade testing process is therefore necessary to cope with these upcoming challenges and will also be a prerequisite to automate steps of testing in this domain. This paper therefore presents an approach to such a standardized testing process.

This paper structures as follows: Section II contains related work, Section III definitions, while Section IV describes some process-internal relations. Section V describes the proposed automotive security testing process and, finally, Section VI concludes this paper.

## II. RELATED WORK

The importance of creating generic testing frameworks or security testbeds to conduct automated security tests in automotive has already been highlighted in the past. More specifically, fuzz testing methods in accordance to industry-specific technologies such as the Controller Area Network (CAN bus) and the vehicle's electronic control unit (ECU) [7], [8], [9], [10] have been created.

Similarly, there are frameworks that address systematic methods of security testing for automotive Bluetooth, Vehicular Ad Hoc Networks (VANETS) in Intelligent Transportation Systems (ITS) and road services [11], [12], [13]. Finally there are integrated security testing frameworks that improve the standardized methods [9], [14], [15], [16], [17]. All of these works, however, do not encompass a defined process for automated security testing of complete automotive systems in a holistic manner. This work therefore complements the standards with a structured testing approach and underpins the technical testing solutions with a structured workflow method. As the upcoming ISO/SAE 21434 [5] is regarded to become the most important guideline, the process aligns to it (see V). There is a supplement to the ISO standard regarding testing (ISO/WD PAS 5112), which, however, is in a larval state.

## III. DEFINITIONS

*Testing* in the context of this paper means verification and validation in the sense of ISO/SAE DIS 21434 (Sections 10.4.2 and 11) [5]. An *Item*, according to the ISO standard mentioned above, is a system or combination thereof to implement a function at the vehicle level. In the sense of this process, an item is understood as a technical concept that defines such a system. A *security goal* is a desired, security-related property of an item, which is analyzed for *threats* and *risks* that lead to *security requirements* that are collected in a *security concept*. A *System-under-test (SUT)* is a concrete technology unit, e.g. a vehicle, a single Electronic Control Unit (ECU) or a software, that concretely instantiates an item that is an examination subject in this process. Therefore, an item can address multiple SUTs(e.g. different cars of the same type). A *test system* is the active unit that carries out the test on an SUT. The proprietor of an SUT or item, respectively, is an *item owner*. The user of a test system is a *test operator*. The compound of SUTs and test system(s) including interfaces and surroundings (e.g. an automotive testbed) constitute a *test environment*. *Test scenarios* in the context of this paper are abstract test descriptions that define what to test by which means, consisting of *test patterns* as their atomic elements that describe single scenario stages. They are derived from a security analysis and requirements definition of the respective item. *Test cases* are the concretization of scenarios for a specific SUT, consisting of *test scripts*, which are executable tests that run on a test system and target an item.

## IV. RELATION BETWEEN TEST SCENARIOS AND TEST CASES

The reason to provide abstract test descriptions, that preempt some test case generation (tcg) operations, is portability to other items. Although test scenarios derive from threat assessments, they are described generic, conveniently in a domain specific language (DSL - e.g. [18]). Test patterns are, consequently, generic means to test a part of the system – e.g. sending a CAN message with a break signal (e.g. *SEND CAN_MSG()*) – that constitute a scenario. These means concretized with test scripts (attack steps - e.g. *./cansend can0 7df#02010d*) that eventually form test cases (concretizations of test scenarios - see Sections V-E and V-F). The difference is that scenarios and patterns only generically describe *what* to do, while the cases and scripts are concrete elaborations *how* to do it. The test cases therefore augment scenarios during the tcg with information (e.g. CAN messages) from a specific SUT in order to test it. The purpose of the test cases is to allow for automated testing using an appropriate framework. Figure 1 illustrates the coherence between the test scenarios with their patterns and the cases with their scripts: scenarios consist of abstract patterns and cases consist of concrete scripts.

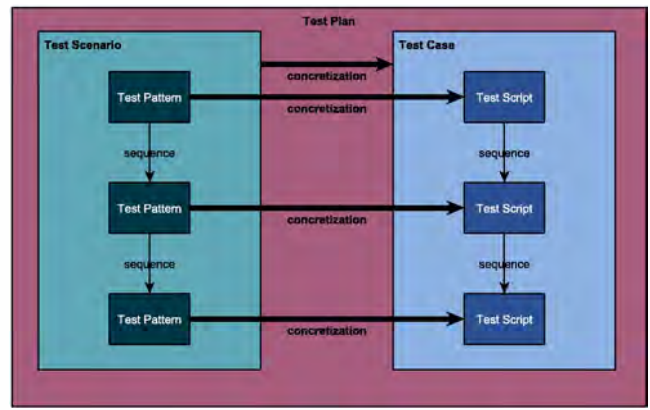

Figure 1. Relation between generic test scenarios/patterns and concrete test cases/scripts

## V. AUTOMOTIVE SECURITY PROCESS

This section outlines the security testing process which consists of the following activities:
1) Define Item;
2) Perform Risk and Threat Analysis;
3) Define Security Concept (testing requirements);
4) Plan Test and Develop Scenarios;
    a) Define Penetration Test Scenarios;
    b) Define Functional and Interface Test Scenarios;
    c) Define Fuzz Testing Scenarios;
    d) Define Vulnerability Scanning Scenarios;
5) Select Test Scripts;
    a) Develop Test Scripts;
    b) Validate Test Scripts;
6) Generate Test Cases;
7) Perform Test;
    a) Prepare Test Environment;
    b) Execute Test Cases
8) Generate Test Reports.

The process is based on the security testing sections of ISO/SAE (DIS) 21434 [5]. The first three activities correspond to respective sections of ISO/SAE DIS 21434 (Sections 9.3 through 9.5) and the selection of test scenario methods is taken from annex (F.2.5). The process further corresponds to the verification part of the standard (10.4.2). Figure 2 provides a process overview and the subsequent sections elaborate descriptions of each process activity. The process is kept generic in order to provide a flexible tool for many purposes, the main use case, however, is security testing of automotive systems. The item definition, risk and threat analysis, and the security concept definition may not be considered as core testing activities, but they provide important information and thus are preconditions to successful security testing.

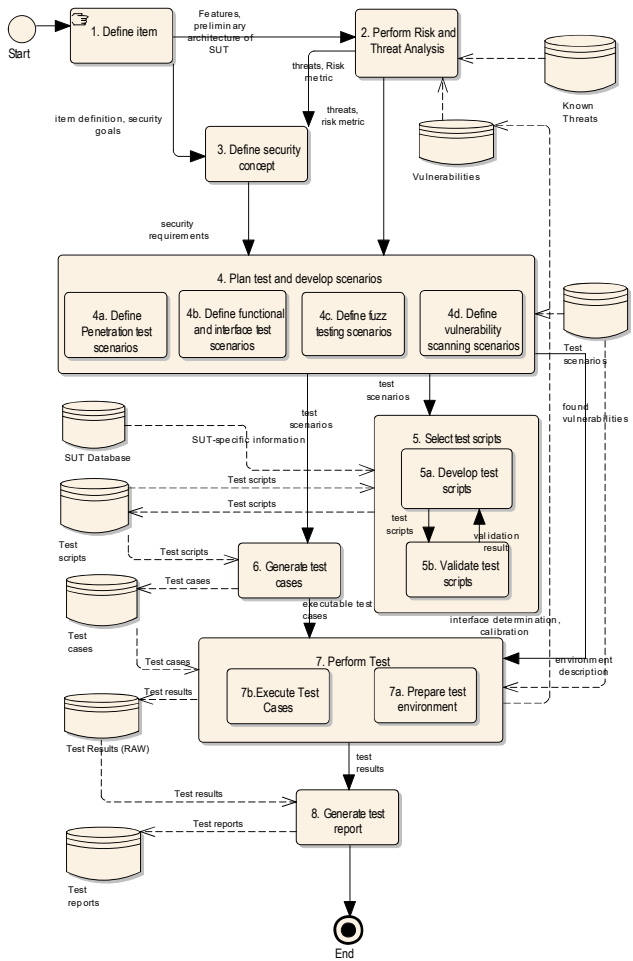

Figure 2. Security Testing Process Layout

### A. Define Item

The item definition is the the first step that collects all relevant information (specification of the components, interfaces and item functions) in a structured format to initiate the testing process and establishes a commonly agreed scope between the item owner and the test operator. The process activity should cover these main aspects [5]:

- The item boundary defining the context of the item;
- The item functions including implementation;
- The preliminary architecture including item internal components and their connections.

The activity output should contain specifications of interfaces, configuration parameters, item boundary and functions, as well as security goals, which serve as an input for the risk and threat assessment (Section V-B). In order to automate this step as far as possible (and allow for black box assessments) the process recommends device fingerprinting [19] and state model derivation [20]. For each defined item (see the process input), the set of tested systems (internal components, interfaces, hardware or software) is defined in the test plan (see V-D) according to the risk assessment. The test result of a concrete SUT is used to improve the testing model for further SUTs of the same item.

### B. Perform Risk and Threat Analysis

A risk and threat analysis (e.g. Threat Analysis and Risk Assessment - TARA [21]) should be performed prior to testing to derive and prioritize requirements and test scenarios. The main goal of a risk and threat analysis is to systematically identify all potential threats to an SUT, and to assess the risk associated with each identified threat in order to determine appropriate risk treatment. Apart from manual expert analysis, (semi-)automated methods are matching models from the item definition with known vulnerabilities or performing automatic model checking [22].

The main steps of such an analysis are [21]:
- Threat Analysis;
- Impact Assessment;
- Risk Assessment (impact x occurrence probability).

The last step also assesses if the risk is acceptable (below a certain threshold). The process does, for the sake of generality, not mandate a specific method but suggests HEAVENS [23] or SAHARA [24]. Finally, a security goal is mapped to each threat to derive security requirements as part of the security concept.

### C. Define Security Concept

Results from the threat analysis (Section V-B) include attack feasibility and countermeasures identified to mitigate specific risks, which derive security requirements that combined yield a security concept. Besides looking for general vulnerabilities and potential issues, the focus of the tests lies on verifying and validating that the defined risk treatment measures are effective. For automated analyses, the requirements to be tested consist of patching the found vulnerability in the model through fixing the underlying faults or additional measures.The requirements are the input for the test planning, as they should be verified with regard to the consistency with the security goals and the item's functionality [5]. The following steps derive such requirements [25]

1) Collect the results from the threat analysis;
2) Define threat countermeasures;
3) Map the resulting threats to countermeasures.

This method allows for using different attack mitigation techniques as building blocks that can later be referred by multiple requirements. The security requirements derive the relevant scenarios for the tests and give direction on what needs to be tested.

*D. Plan Test and Develop Scenarios*

A security test plan should organize the security testing process and contain the following elements [26]:
- Purpose/Objectives;
- SuT overview and Test scope;
- Risk analysis;
- Test strategy and requirements;
- Test environment (Hardware- or Software-in-the-Loop, actual vehicle, etc.);
- Test case specifications;
- Test execution and termination criteria.

This also corresponds to a verification and a validation specification according to ISO/SAE DIS 21434 (10.5 and 11.5, respectively) [5]. The output of this activity is a set of defined test scenarios which, dependent on the risk level and attack feasibility, apply different techniques. Test scenarios are between system requirements and test cases [27] and are abstract test descriptions (consisting of test patterns) that define which vulnerabilities and requirements are specifically tested with which methods. Testing methods, based on the identified risks and threats, are [5]:
- Functional testing
- Interface testing
- Static code analysis;
- Penetration testing;
- Vulnerability scanning;
- Fuzz testing.

A Test pattern is the generic description of a single step inside a test (normally an action during an attack) including potentially used tools, but not specific to an SuT. These methods should be made concrete with test scripts (containing e.g. an exploit) that eventually form test cases (see Section IV for the coherence between test scenarios and test cases). Scenarios are derived by requirements analysis, equivalences classes, boundary values analysis and error guessing [5].Furthermore, attack patterns are derived from generalizing existing attacks that may be derived from open databases (for well-known attacks) or intrusion detection signatures as well as actual attack analyses. For an automated test system implementing the process, the availability of attacks for the derived models (or constituting components, respectively) determines the used test patterns and, thus, the test plan, including the test methods.

*1) Define Penetration Test Scenarios:* Penetration testing is the legal and authorized process of exploiting systems in order to retrieve information which is important for enhancing security of the system. Penetration tests focus on specific aspects of security and are deployed manually or semi-automatic. To extend the capabilities, global-based adversarial activities must be deployed to maintain a holistic view of the system and deploy security tests from the adversary's perspective. The above methods are called red team assessments which usually include penetration tests; however, such methods extend the whole process [28]. A successful penetration testing methodology will discover functional weaknesses, design flaws and provide recommendations for security improvement [29]. To deploy penetration test scenarios, the scope and the context for deployment of appropriate attack strategies with respect to the system's potential weaknesses must be defined. In penetration testing, it is possible to attack vehicles without in-depth knowledge (black box) or from the inside (white box - meaning that some or full information is available to the red team). The process suggests cyber kill chain [30] and attack trees [31], where the latter approach allows for automated decision making for generating attack vectors.

*2) Define Functional and Interface Security Testing Scenarios:* Functional tests assess the system's adherence to its functional requirements (correctness)and take place throughout the whole process and at different levels of abstraction. Testing security functions focuses on testing the security requirements. Typical security requirements may include specific elements of confidentiality, integrity, authentication, availability, authorization and non-repudiation. There are two possibilities of formulating security requirements: 1. positive requirements and 2. negative requirements. Positive formulated requirements describe how a security function should work. Negative requirements state behaviour that the software should not exhibit. The mapping of requirements to specific software artifacts could be problematic for such requirements, since this kind of requirement is not implemented in a specific place[26]. When negative requirements are tested, security testers look for common mistakes and test assumed weaknesses in the application. The emphasis is on finding vulnerabilities, often by executing misuse tests. To derive the test cases, the following steps need to be carried out:

1) Identify functions expected to perform.
2) Create test cases based on the function methods.
3) Determine the output based on the function specifications.

*3) Define Fuzzing Scenarios:* Fuzzing is a technique to use random input in order to put an SUT into a non-intended state to uncover errors, which could be more efficient than structured testing [32]. However, randomness of fuzz testing does not have to be complete but adapted to an SUT using passive listening [33]. A fuzzer consists of a generator (combining valid and random parts), a delivery mechanism, a monitoring system and a test oracle [34]. The oracle, that determines the test result (i. e. pass/fail), is obtained by monitoring communications or using specific protocols (like XCP) as well [35]. Using fuzzing techniques, it is possible to attack automobiles without any in-depth knowledge [36]. In principle, any component that shows an external interface can be fuzzed.

Fuzz testing can [37]:
- be used to reverse engineer vehicle messages;
- be used to disrupt vehicle's communication network;
- be a form of cyber attack;
- lead to vehicle component damage.

For a significantly large test space, fuzzing should be combined with combinatorics to select test cases and be run in parallel as long as a test series runs and the space is not covered.

*4) Vulnerability Scanning Scenarios:* Vulnerability scanning uses tools, called vulnerability scanners, that compare a vulnerability database with the information obtained from a network scan to find possible vulnerabilities in the network [38].

A scanner typically enumerates known software vulnerabilities and provide a comprehensive baseline of existing vulnerabilities. To perform effective vulnerability scanning, the tools should be selected based on the scanning scope. This scope is needed to define and create the vulnerability scanning scenarios. A typical scenario for using vulnerability scanning is:
- Define which system to scan (i.e. the SUT or components thereof);
- Define the tool that should be used for the scanning;
- Perform the scan;
- Analyze the resulting report (i.e. identify relevant vulnerabilities);
- Specify further analysis/testing tasks.

For automation, the results (in machine readable form) serve as an input for other scenarios.

*E. Select Test Scripts*

This section concerns the transition of generic test descriptions (from the test scenarios) addressing vulnerabilities (found in the threat assessment) into concrete test scripts to be executed onto a specific SUT. Test scripts are selected from a database, if available, or otherwise developed.

*1) Develop Test Scripts:* The purpose of this activity, in general, is to populate a test script database with relevant tests, particularly attacks. The scripts correspond to the plan and implement defined test patterns. This activity is optional and carried out if no appropriate test script is present beforehand. The scripts are concrete implementations of test patterns, making use of the tools outlined in the scenario description targeting towards a specific SUT. A test script is an executable script that contains:
- The testing tool(s) to be used (parameters, interfaces, oracle may be derived from the test scenario;
- Needed parameters and information specific to the SUT;
- Specifics of the test system (e.g. using Linux, availability of a certain compiler/interpreter, etc.).

Similar to test scenarios, attack scripts are derived from open sources by observing actual attacks. Test scripts are created by analysing an SUT or they are derived from various structured approaches like attack trees [39]. To ensure that the case generation step (Section V-F) can re-use scripts from the database, the current step should:
1) Either match an existing script;
2) Or develop a test script to the specifications of the test scenario.

In the latter case, extensive technical knowledge about the SUT or further specifics might be needed (e.g., an interpretation file for particular CAN messages).

*2) Validate Test Scripts:* This optional activity applies to newly developed test scripts to validate them before actual tests. New test scripts are tried out on simulated or actual SUTs in a simulated or actual test environment. Expected outcomes (derived from the test oracle) are compared to actually acquired results. SUTs should be chosen in a way that both positive and negative results can be obtained in specific well-defined conditions. In order to validate a test script, the environment should fulfill any prerequisite set in the test script (similar to an actual test). Similar to test cases, the validation of the test script should contain different SUTs or configurations thereof that include:
1) an SUT configuration with a successful attack (positive validation);
2) an SUT configuration with an unsuccessful attack (negative validation);
3) several edge cases.

The validation test coverage should be comparable to coverages in actual tests (see Section V-F).

*F. Generate Test Cases*

A test case includes multiple items from this non-exhaustive list derived and extended from [40], [41], [42]:
- Test purpose and objectives;
- SUT/Function description (including software/hardware/firmware configurations);
- Environmental needs including dependencies;
- Procedural requirements, test setup and condition;
- Test activities and input data;
- Expected results, completion, stopping and resumption criteria (Pass/Fail criteria including metrics);

- Traceability to related requirements and threats;
- Variability and quality attributes.

In the context of this process, the test case generation (tcg) is the fusion of a generic test scenario (Section V-D) and the test scripts (Section V-E) that are specific to a distinct SUT. Augmenting the scenarios with specific information from an SUT database translates the test scenarios into executable test cases. With both parts available in machine-readable form, this activity is easy to automate. Combinatorial testing [43] allows for an efficient coverage/effort ratio. The tcg can re-use the threat modelling outcomes in conjunction with a test script database, giving the opportunity of automating the process using a framework (e.g. [44]) If a clear model is lacking completely, test coverage is most important.

### G. Perform Testing

To execute the test, a test environment shall be established using an description from the scenarios. The environment description contains all required prerequisites, while the test cases contain the performed operations

*1) Prepare Test Environment:* Two inputs are needed to prepare a test environment: (a) an environment configuration and (b) interface descriptions. The resulting test environment template is then used to execute tests. A configuration consists primarily of the system under test (SUT) and applicable test categories, including system and service preconditions. Interface descriptions contain their stimulation and provisions, as well as verification procedures for their claimed properties. For automation, they are organized in an object-oriented, serializable manner. The resulting combination of the environment and the interface description form a test environment template to applied with different test cases from diverse categories, ideally in a microservices-based, containerized style. This activity also includes saving a pre-attack state of the SUT and a clean-up procedure after the conducted test (e.g. if a test involves flashing ECUs).

*2) Execute Test Cases:* Each test case consists of a sequence of test scripts (as minimum verifiable actions – MVAs) that can be combined to form more complex sequences. Resulting sequences can be combined again. The combinations can also contain permutations and re-organizations of scripts. For automation, final commands take a shell-executable form. Test cases create specific outputs on defined interfaces. This output is consumed by an interface module that transforms the output into a correct call for the associated physical interface and the returned response. A completed test case output is subsequently converted into a standardized test result. Test results are stored and used as input for other test cases, further analysis, and reporting (see V-H), including relevant meta data in a standardized format.

### H. Generate Test Report

A test report is a presentation of the combined results of the process, it should contain:
- A management summary;
- An SUT description;
- start time and duration;
- An aggregated overview (dashboard);
- The approach/method used;
- Findings (passed and failed tests);

For the executed tests, pass and fail information (and in case of failed tests: sufficient information to understand the problem) must be given. In both cases, reference links to goals, requirements, used tools, the raw data, the test results (including risk levels and severity categorization and conflicts with regulations, policies or best practices) and information about aspects that were not tested (not planned, technical problems, lack of time, funds or tools, etc.) need to be included. The testing report should also correspond to a verification and a validation report according to ISO/SAE DIS 21434 (10.5 and 11.5) [5].

## VI. Conclusion and Outlook

This paper outlined a process for testing the cybersecurity of (particularly automotive) systems to fill the gap between existing standards for automotive security engineering and their hands-on, actual-system testing. The process provides a comprehensive, automatable approach for system testing based on ISO/SAE DIS 21343. Due to rising complexity and regulators' requirements this is necessary as it facilitates a conceivable need for industrializing automotive cybersecurity testing. The process itself is arranged generically in order to allow for using already existing procedures (e.g. a present risk assessment process) not mandating any specific method. Future work will therefore involve a reference implementation on both processual and technical level.

## Acknowledgement

This work was supported by the H2020-ECSEL program of the European Commission; grant no. 783119, SE-CREDAS project. Special thanks to Rosita Jupri, Behrooz Sangcholie and Rauli Kaksonen for their help.